# Influence of drying temperature on morphology of MAPbI$_3$ thin films and the performance of solar cells

Hao Zhang [1], Yalan Wang [1], Hong Wang [1], Meryang Ma [1], Shuai Dong [1,*], and Qingyu Xu [1,2,*]

[1]School of Physics, Southeast University, Nanjing 211189, China

[2]National Laboratory of Solid State Microstructures, Nanjing University, Nanjing 210093 China

*E-mail: sdong@seu.edu.cn (S. D.); xuqingyu@seu.edu.cn (Q. X.)

## Abstract

Photoelectric conversion efficiency of organic-inorganic perovskite solar cells has been rapidly raised and attracted great attention in recent years. The quality of perovskite films is vital for the performance of devices. We used the anti-solvent method to prepare CH$_3$NH$_3$PbI$_3$ thin films by spin coating and dried them at various temperature to transform adduct MAI · PbI$_2$ · DMSO into CH$_3$NH$_3$PbI$_3$. We researched in detail on the relationship between surface morphology of MAPbI3 thin films fabricated by the anti-solvent method and various drying temperature. We found that surface roughness and grain size of CH$_3$NH$_3$PbI$_3$ films together increased with increasing drying temperature. The larger grain size could efficiently reduce crystal boundaries which is advantageous for the suppression of photo-induced charge carrier recombination resulting in increase of FF. However, increase of surface roughness resulted in larger contact area at interface which might produce more tarp states and poorer wettability of HTM solution leading in decrease of J$_{sc}$. Surface morphology of MAPbI3 layer on the performance of solar cell devices is also an important research issue. By optimizing the drying temperature to 60 °C, the highest efficiency of 14.4% was achieved for the CH$_3$NH$_3$PbI$_3$-based solar cell devices.

## 1. Introduction

To meet the growth of energy demand and reduce environmental pollution in modern society, renewable energy sources and technologies have attracted much attention and continue to be developed. Among all natural energy sources, solar energy is the most abundant which can provide over 10000 times of power required by the entire planet.[1] In recent years, organic-inorganic hybrid halide perovskite solar cells become a hotspot due to their appropriate band gap, high absorption coefficient, long charge diffusion length and solution processability.[2] The formula of these perovskite compounds is ABX$_3$ (A cations are organic like CH$_3$NH$^{3+}$, C$_2$H$_5$NH$^{3+}$ and HC(NH$_2$)$^{2+}$; B cations are metal elements such as Pb$^{2+}$ and Sn$^{2+}$; X anions are halogen elements (I$^-$, Br$^-$, Cl$^-$) ).[3,4] The organic-inorganic hybrid halide perovskite material (CH$_3$NH$_3$PbBr$_3$) with an efficiency of 2.2% in 2006 and

$CH_3NH_3PbI_3$ with an efficiency of 3.8% in 2009 were first reported, which were used in photovoltaics as a sensitizer replacing the dye pigment in dye sensitized solar cells (DSSCs) by Miyasaka and co-workers.[5,6] After then, the photoelectric conversion efficiency of perovskite solar cells increased rapidly up to a record of 22.1% in early 2016, but this value is still far from the theoretical efficiency limit.[7,8]

Traditional mesoporous organic-inorganic hybrid halide perovskite solar cells are mainly composed of a bottom electrode (fluorine doped tin oxide), an electron transporting layer made of n-type semiconductor ($TiO_2$, $SnO_2$, ZnO and so on), a mesoporous layer made of n-type semiconductor nanoparticles such as $TiO_2$ and ZnO, or insulating nanoparticles such as $Al_2O_3$ and $ZrO_2$, a light absorb layer, an hole transport layer (HTL) made of p-type semiconductor (Spiro-OMeTAD, CuSCN, NiO and so on) and a top electrode (Ag, Au, Al, C and so on).[4,9–11] For this kind of solar cells, quantity of $MAPbI_3$ thin films such as grain size, surface morphology and so on are the key parameters to be optimized to obtain high performance of devices.

Anti-solvent method was adopted to prepare $MAPbI_3$ thin films which washed out DMF solution in thin films with nonpolar diethyl ether solution and produced the adduct MAI·$PbI_2$·DMSO due to function of Lewis base and acid.[12,13] Then, drying the substrates for minutes can fully transform the adduct MAI·$PbI_2$·DMSO into $MAPbI_3$ without appearance of pinholes in $MAPbI_3$ thin films. Now, the anti-solvent method has become one of the most popular selection in all one-step spin-coating methods. So far, there has not been many works focus in detail on the effect of drying temperature on $MAPbI_3$ thin films fabricated by the anti-solvent method rather than common spin-coating methods. Furthermore, these researches did not explain the influence change of morphology on performance of devices in detail.[14,15] In our work, we detailedly researched on the morphology of $MAPbI_3$ thin films such as grain size and surface roughness caused by change of the drying temperature and the relationship between the change of morphology and performance of devices. As a result, according to EIS and steady-state PL measurement, we found that increasing of drying temperature of $MAPbI_3$ thin films fabricated by the anti-solvent method had advantageous and disadvantageous influence on devices and there existed an optima value can balance the two aspect.

## 2. Materials and Methods

### 2.1. Materials

Glass coated Fluorine-doped tin oxide (7Ω/sq) was available from Nippon Sheet Glass Company. Mesoporous $TiO_2$ paste (18-NRT) was acquired from Australia Dyesol Co. Tetrabutyl titanate (99%), absolute ethyl alcohol (99.7%), methylamine (30wt%-33wt% in absolute ethyl alcohol), chlorobenzene (99.5%) and lead iodide (99.9%) were obtained commercially from Aladdin Company. N, N-Dimethylformamide (99.5%), Dimethyl sulfoxide (99%), acetonitrile (99.8%), diethyl ether (99.5%) and Hydroiodic acid (45wt% in water) were bought from Sinopharm Chemical Reagent Co. Ltd.. Spiro-OMeTAD (99%) was supplied by Feiming technology Co. Ltd. Lithium bis(trifluoromethanesulfonyl)imide (Li-TFSI) (99%) and tributyl phosphate (TBP) (99%) were purchased from Xi'an Polymer

Light Technology Co. Cobalt bis (trifl uoromethanesulfonyl) imide (Co-TFSI) (99%) was purchased from Shanghai Materwin new materials Co. Ltd. Argent grain (99.999%) was purchased by Beijing PuRui new materials Co..

2.2. Methods

Firstly, a FTO substrate of right size was incised and then cleaned by ultrasonic cleaning in the detergent water, deionized water and absolute ethyl alcohol, respectively. Tetrabutyl titanate ethyl alcohol solution was spin coated on the substrate at 5000 rpm for 30 seconds to prepare 80 nm compact $TiO_2$ layer that was n-type semiconductor and annealed for 40 minutes at 450 °C. 200 nm thick mesoporous $TiO_2$ layer was prepared by spin coating mesoporous $TiO_2$ diluent (mesoporous $TiO_2$ paste: ethyl alcohol = 1 : 4) at 4000 rpm for 30 seconds and then was heated at 450 °C for 60 minutes.

$CH_3NH_3I$ was synthesized with hydroiodic acid, methylamine and absolute ethyl alcohol.[16] $CH_3NH_3PbI_3$ solution was synthesized by mixing 0.795 g MAI, 2.305 g $PbI_2$, 3 g DMF and 400 μL DMSO together and magnetic stirring for 4 hours. A 300 nm thick $MAPbI_3$ thin film was fabricated by spin coating $MAPbI_3$ solution on the substrate at 4000 rpm for 30 seconds. In the process of spin coating, 1 ml diethyl ether was dropped on the substrate to wash out DMF. When the process of spin coating ended, the substrate was heated at various drying temperatures from 40 °C to 80 °C to evaporate the remaining DMSO until the thin film became red brown and transform adduct MAI·$PbI_2$·DMSO into $MAPbI_3$. Afterwards, the substrate was annealed at 100 °C for 2 minutes.

Around 250 nm thick HTL was prepared on the $MAPbI_3$ layer by spin-coating the hole transport material (HTM) solution mixed of 288 μL TBP, 175 μL Li-TFSI solution (520 mg Li-TSFI in 1 ml acetonitrile), 290 μL Co-TFSI solution (300 mg Co-TSFI in 1 ml acetonitrile) and 20 mL chlorobenzene at 5000 rpm for 30 seconds and placed in a glove box for one night for drying. Finally, 80 nm thick Ag layer as top electrode was deposited by thermal evaporation.

2.3. Measurements

In this experiment, X-ray diffraction (XRD) patterns were measured by an X-ray diffractometer (Rigaku Smartlab3) which used Cu Kα as the radiation source to analyze the structure and constituents in film. Images of cross-section and surfaces of samples were taken by a scanning electron microscope (FEI Inspect F50). Steady-state photoluminescence spectrums were obtained by a fluorescence spectrometer made by Horiba Jobin Yvon using the laser of 440 nm wavelength. Optical absorption spectra were measured by an ultraviolet and visible spectrophotometer which was manufactured by Hitachi Co. Atomic force microscope (AFM) images were taken by a BioScope Resolve[TM]. Current-voltage (I-V) curves was measured by Keithley 2400 under Newport Oriel 91.192 simulated illumination (AM1.5, 100 mw/$cm^2$). External quantum efficiency (EQE) were measured by an EQE test system fabricated by Newport Co.. Electrochemical impedance spectra (EIS) was measured by an electrochemical workstation produced by Shanghai ChenHua instruments Co..

# 3. Results and discussion

The experimental process of anti-solvent method to prepare MAPbI$_3$ thin films is shown in figure 1(a). In the process of spin coating, nonpolar diethyl ether solution was dropped to wash away DMF in the MAPbI$_3$ thin films. Adduct MAI • PbI$_2$ • DMSO was produced and the thin film appear slightly transparent yellow. A drying process was utilized to remove DMSO and transform the adduct MAI·PbI$_2$·DMSO into pure MAPbI$_3$ that made the thin film's color change into red brown. We dried the MAI • PbI$_2$ • DMSO thin film at various temperature (40 °C, 50 °C, 60 °C, 70 °C and 80 °C) and study the influence of different drying temperature on the perovskite thin film and the performance of whole device. XRD patterns of the MAPbI$_3$ thin film prepared by this method at the drying temperature of 60 °C before and after annealing are displayed in figure 1(b)-(c). It can be seen that pure MAPbI$_3$ can be obtained after drying and the relative intensity of the {002} is observably enhanced after annealing.[17] Similar phenomena have been observed in all the MAPbI$_3$ thin films dried at other temperatures, as shown in figure S1. PL and UV-vis absorption spectra (figure 1(c)) reveals the PL peak at 778 nm, indicating that the bandgap of corresponding MAPbI$_3$ thin films is around 1.59 eV and absorption onset is at 750 nm with large absorption intensity under 540 nm.

SEM images of MAPbI$_3$ thin films' surfaces at different drying temperature are shown in figure 2. From figure 2, we can see that MAPbI$_3$ grain size after annealing is almost unchanged when drying temperature is under 60 °C. It is obvious that MAPbI$_3$ grain size become gradual larger with the increasing drying temperature. Distribution of MAPbI$_3$ grain size according to measurement tools based on SEM images Figure 2 can be seen in figure 3(a) – (e). It is obvious that the range of grain size is mainly from 50 nm to 650 nm. Proportion of 150 nm – 450 nm is predominant and proportion of 450 nm – 650 nm increases with increasing drying temperature. When drying temperature is up to 80 °C, grain size in the range of 650 nm – 750 nm appears and the range of 50 nm – 150 nm disappears. In a word, proportion of larger grain size is becoming more with the increasing of drying temperature. Relationship between range of grain size and drying temperature is shown in figure 3(f). We can find a trend that grain size sharply increases with increasing drying temperature above 60 °C, and slightly rises with drying temperature is below 60 °C. As we all know, increasing annealing temperature can contribute to growth of MAPbI$_3$ crystalline grain.[18] So, drying temperature has similar effect like annealing temperature on MAPbI$_3$ crystalline grain. There are trap states in crystallite surfaces and interfaces where the bulk crystalline symmetry is broken resulting in the recombination of charge carriers.[19] Larger grain size can reduce grain boundary which decreases trap states and suppresses charge recombination and makes for transporting of electron into the TiO$_2$ layer in the MAPbI$_3$ thin film.[20] The steady-state PL spectra of perovskite thin films on TiO$_2$ layer fabricated at the drying temperature of 40 °C, 50 °C, 60 °C, 70 °C and 80 °C are displayed in figure 4. From this image, there is an evident reduction of PL intensity with increasing of drying temperature. This can be understood by increasing of grain size which results in decrease of grain boundaries with carrier trapping sites which makes for increasing of carrier transport mobility in MAPbI$_3$ thin films and efficient electron injection to the TiO$_2$ layer from perovskite layer.[21–23]

We find that the time for fully transformation from adduct MAI • PbI$_2$ • DMSO (colorless transparent) into MAPbI$_3$ (red brown) is shorter with increasing of drying temperature (figure 5(a)). Higher drying temperature accelerates the transformation into MAPbI$_3$ and evaporation of DMSO causing more rapid change of thin films' color. Surface topography of MAPbI$_3$ thin films fabricated at drying temperature of 40 °C, 50 °C, 60 °C, 70 °C and 80 °C was surveyed by AFM (Figure 5(b) – (f)). According to these AFM images, root mean square (RMS) roughness of MAPbI$_3$ thin films can be obtained based on these AFM images. Relationship between RMS roughness and drying temperature is shown in figure 5(g). Based on Figure 1(a), we can see that there is an apparent increase of average MAPbI$_3$ grain size with increasing the drying temperature we used from 60 °C to 70 °C. We suppose that this is due to the aggregation of small adjacent MAPbI$_3$ grains just crystallized from intermediate product (MAI-DMSO-PbI$_2$), which becomes more and more obvious with drying temperature above 60 °C in drying process.[18,24] Figure 1(b) shows that RMS roughness of MAPbI$_3$ thin films rises abruptly with increasing drying temperature above 60 °C, in accordance with variation trend of grain size. In addition, we speculate that a faster evaporation of DMSO may slightly contribute to increase of surface roughness of MAPbI$_3$ thin films. A smoother MAPbI$_3$ thin films can effectively contribute to better contact with its capping layer which is advantageous for the performance of devices.[25] Charge carrier trapping at the interface between the MAPbI$_3$ thin film and the HTL can influence transport of charge carrier.[26] Therefore, increase of roughness of MAPbI$_3$ thin films surfaces can slightly enlarge interface area with the HTL which can increase the number of trap states at the interface and reduce short-circuit current density resulting in poorer device performance.

For the further research on effect of interface roughness of MAPbI$_3$ thin films on performance of devices, electrochemical impedance spectroscopy (EIS) measurements were conducted under dark, as shown in figure 6. The inset in figure 6 is the equivalent circuit used for fitting the EIS. The series resistance $R_s$ is the resistance of the conductive substrate and the wire electrode, $R_{sc}$ is the contact resistance related to the charge transfer resistance between the perovskite and selective contacts and transport resistance in the HTL and the electron transport layer. $R_{rec}$ is the recombination resistance which is influenced by the recombination rate related to perovskite layer.[27–31] In the Nyquist plots, the arc in high frequency (left part) and in the low frequency (right part) are related to the transport of hole in the hole transport layer of devices and recombination resistance ($R_{rec}$) respectively.[32,33] It is obvious as shown in figure 6 that the radius of arcs at low frequency increases with the increase of drying temperature. Lowering of radius indicates the decrease of $R_{rec}$ to avoid the carrier recombination.[29] So, we can know that recombination of charge carrier inside the MAPbI$_3$ thin film gradually decreases with increase of drying temperature due to increase of MAPbI$_3$ grain size in excellence accord with steady-state PL showed in figure 4. Figure S2 shows the relationship between contact angles of HTM solution and surface roughness of MAPbI$_3$ thin films. It is obvious that contact angles of HTM solution increases with increase of surface roughness. So, increasing of surface roughness can lead in contact problems such as poorer wettability of HTM solution and increase of contact resistance. Nevertheless, increase of surface roughness can also add contact areas which makes for the transporting of hole into HTL.

Solar cells were fabricated with different MAPbI$_3$ thin films prepared at drying temperature of 40 °C, 50 °C, 60 °C, 70 °C and 80 °C, respectively. The cross-sectional SEM image of perovskite solar cells fabricated in this work are displayed in figure 7(a) which shows us the whole structure of devices. We fabricated 15 perovskite solar cells by the anti-solvent method with drying temperature of 40 °C, 50 °C, 60 °C, 70 °C and 80 °C respectively. Then, we measured the open-circuit voltage ($V_{oc}$), the short-circuit current density ($J_{sc}$), the fill factor (FF) and the photo-to-current efficiency (PCE) of them showed in figure 7 (b) - (e). To make sure the reproducibility and reliability of data we obtained, we used the same appliances, chemical materials and fabrication methods, except for the annealing temperature of MAPbI3 thin films. Among these images, round dots means average values, top and bottom lines represent maximum and minimum values respectively, middle lines show median, boxes contain intermediate 50% values of statistical data. From figure 7(b), we can know that the average $V_{oc}$ of devices dried at various temperature are fairly near within 0.85 V - 0.9 V which indicates that the change of drying temperature has no influence on $V_{oc}$ of devices. According to figure 7(c), it is obvious that the $J_{sc}$ range of devices sharply enlarges with drying temperature increasing from 50 °C to 60 °C and slightly decreases above 60 °C. Figure 7(d) displays that the FF range of devices gradually enlarges with the increasing of drying temperature in agreement with the variation trend of MAPbI$_3$ grain size. Enlarging grain size reduces grain boundary resulting in decreasing of charge carrier recombination which add FF of devices.[34] We can see from Figure 7(e) that the PCE range of devices slightly increases with increasing drying temperature below 50 °C and obviously increases when the drying temperature rises from 50 °C to 60 °C. With the drying temperature increasing above 60 °C, average PCE of devices are slightly get decreased. Therefore, we can conclude that the drying temperature advantageous for the performance of devices is 60 °C.

Finally, the highest photoelectric conversion efficiency of 14.4% could be obtained when we selects 60 °C as drying temperature, and the corresponding J-V curve and EQE are shown in figure 7(f) and Figure S3. The $J_{sc}$ is 23.1 mA/cm$^2$, the $V_{oc}$ is 0.997 V, FF is 62.55 %. Poorer efficiencies would be obtained if we chose higher or lower drying temperatures. The mechanism can be understood as following: Grain size of MAPbI$_3$ layer gradually increases with increasing drying temperature resulting in reduction of grain boundary and recombination rate of charge carrier, meanwhile surface roughness of MAPbI$_3$ layer gradually increases with increasing drying temperature below 60 °C and almost keeps stable above 70 °C. Increase of surface roughness of the interface structure between MAPbI$_3$ and HTM layers can cause more charge carrier trapping and contact resistance. For exemplifying this issue, we fabricated the solar cells with different surface roughness of MAPbI$_3$ layers using vapor in higher temperature, and we found that the performance of devices decrease with increasing surface roughness of MAPbI$_3$ layers (Supporting materials, figure S2).[35] Therefore, with the increasing drying temperature below 70 °C, contact problem such as the change of wettability between HTM solution and surface of MAPbI$_3$ thin films can be caused due to increasing of surface roughness. Furthermore, the contact area between MAPbI$_3$ and HTM layers increases, which is beneficial for the collecting of holes from MAPbI$_3$ layer to HTM layer. Drying temperature of 60 °C is the optimum value, which can reach the balance between all advantageous and disadvantageous aspects caused by the change of morphology so that the highest efficiency of 14.4% can be obtained. Our results suggest that the surface roughness of each layer in the organic-inorganic perovskite solar cells prepared by the solution processed technique is also a significant parameter which should be considered, since the wettability of solution may significantly influence the interface structure quality and roughness can add the contact resistance of devices.

## 4. Discussion and Conclusions

In this work, we studied on relationship between morphology such as change of interface roughness and $MAPbI_3$ grain size caused by selected various drying temperature after the process of spin coating and performance of perovskite solar cells fabricated by the anti-solvent method in detail. Increasing drying temperature could result in growth of $MAPbI_3$ grains which are beneficial for reduction of recombination of photo-induced electrons and holes in $MAPbI_3$ layer resulting in the gradual increase of FF. However, it will also cause surfaces of $MAPbI_3$ thin films rougher, resulting in increase of charge carrier trapping sites at interface between $MAPbI_3$ film and HTM, which is disadvantageous for injection of hole from perovskite layer into HTM and device performance which can lead in decrease of $J_{sc}$. Increasing surface roughness can also cause worse wettability of HTM solution on the $MAPbI_3$ layers. By compromising these advantageous and disadvantageous factors influencing performance of devices together, optimized drying temperature of 60 $^o$C was selected, and the highest efficiency 14.4% of perovskite solar cells has been obtained.

## Acknowledgments

This work is supported by the National Natural Science Foundation of China (51471085, 11674055), the Natural Science Foundation of Jiangsu Province of China (BK20151400), and the open research fund of Key Laboratory of MEMS of Ministry of Education, Southeast University.

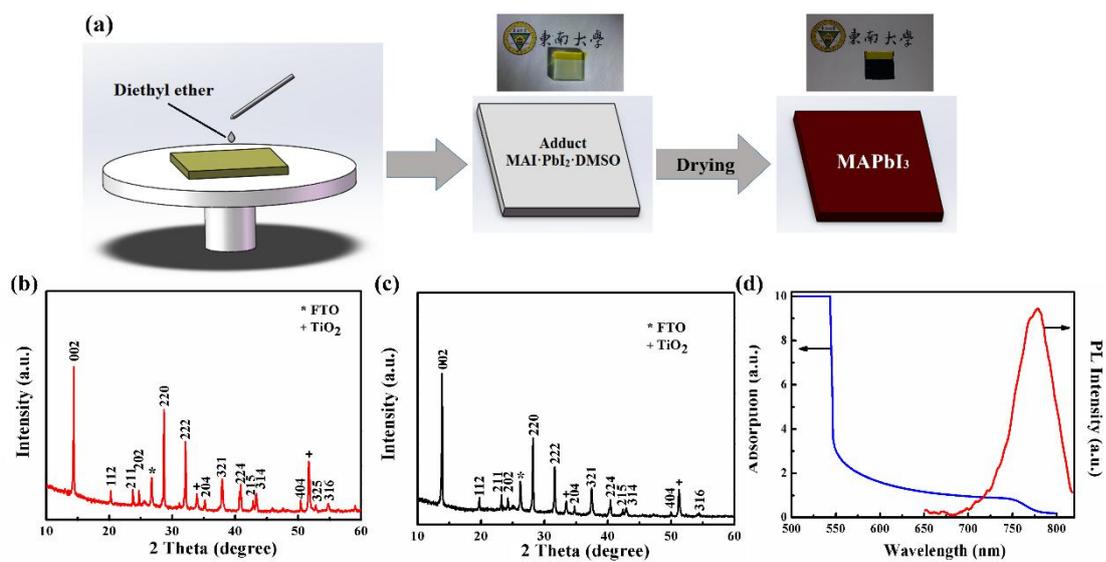

**Figure 1.** (a) Schematic representation of the anti-solvent method to fabricate MAPbI$_3$ thin films (with photos of MAPbI$_3$ thin film before and after drying). XRD patterns of MAPbI$_3$ thin film prepared by anti-solvent method at drying temperature of 60 °C (b) before and (c) after 100 °C annealing. (d) The steady-state PL spectra and UV-vis absorption spectra of the MAPbI$_3$ thin film.

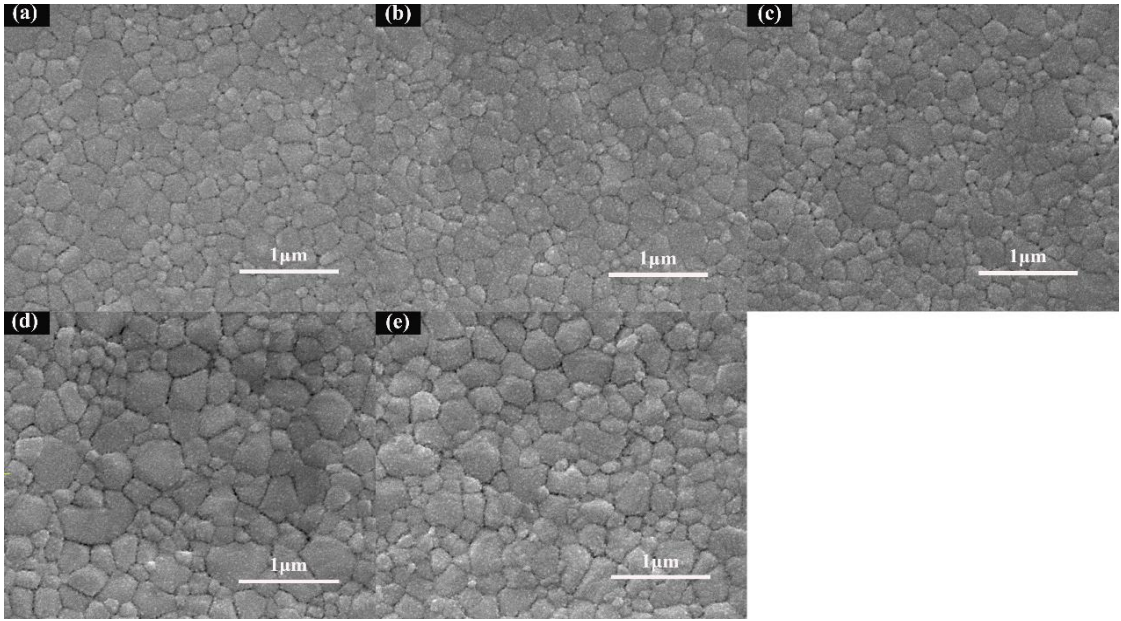

**Figure 2.** SEM images of MAPbI$_3$ thin films' surfaces at drying temperature of 40 ºC, 50ºC, 60 ºC, 70 ºC and 80 ºC after annealing at 100 ºC.

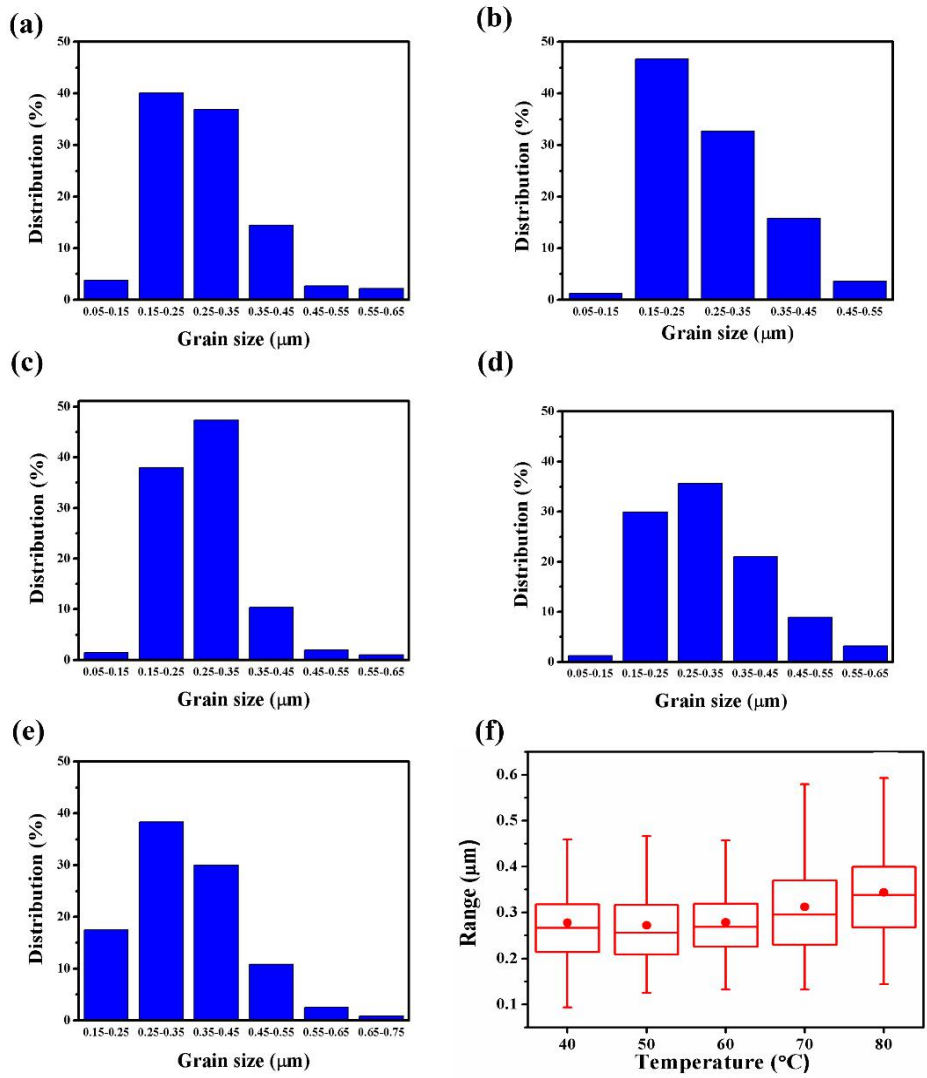

**Figure 3.** Distribution of grain size at the drying temperature of (a) 40 °C, (b) 50 °C, (c) 60 °C, (d) 70 °C and (e) 80 °C based on SEM images. (f) Grain size range of MAPbI$_3$ thin films at the drying temperature of 40 °C, 50 °C, 60 °C, 70 °C and 80 °C.

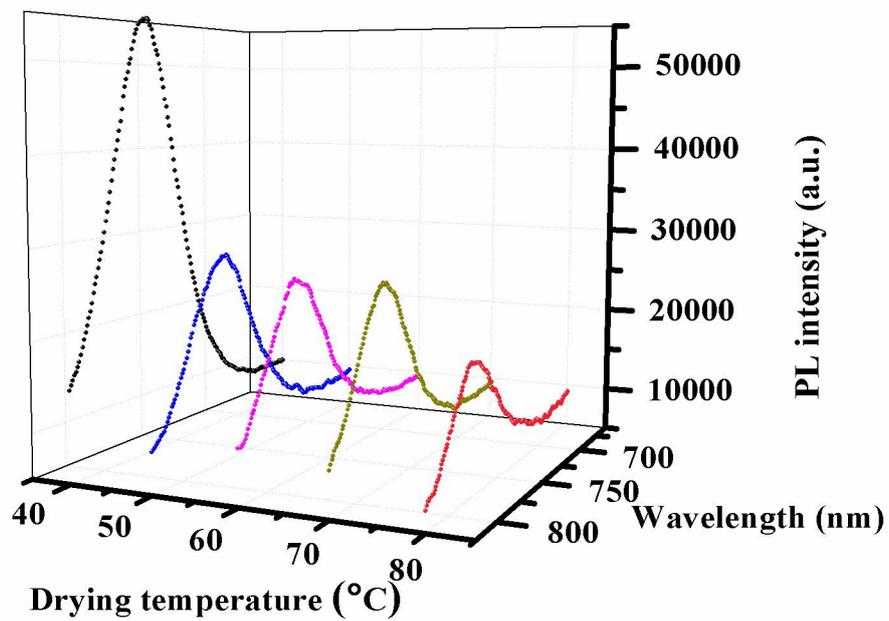

**Figure 4.** Steady-state PL spectra of perovskite thin films fabricated by the anti-solvent method at the various drying temperature.

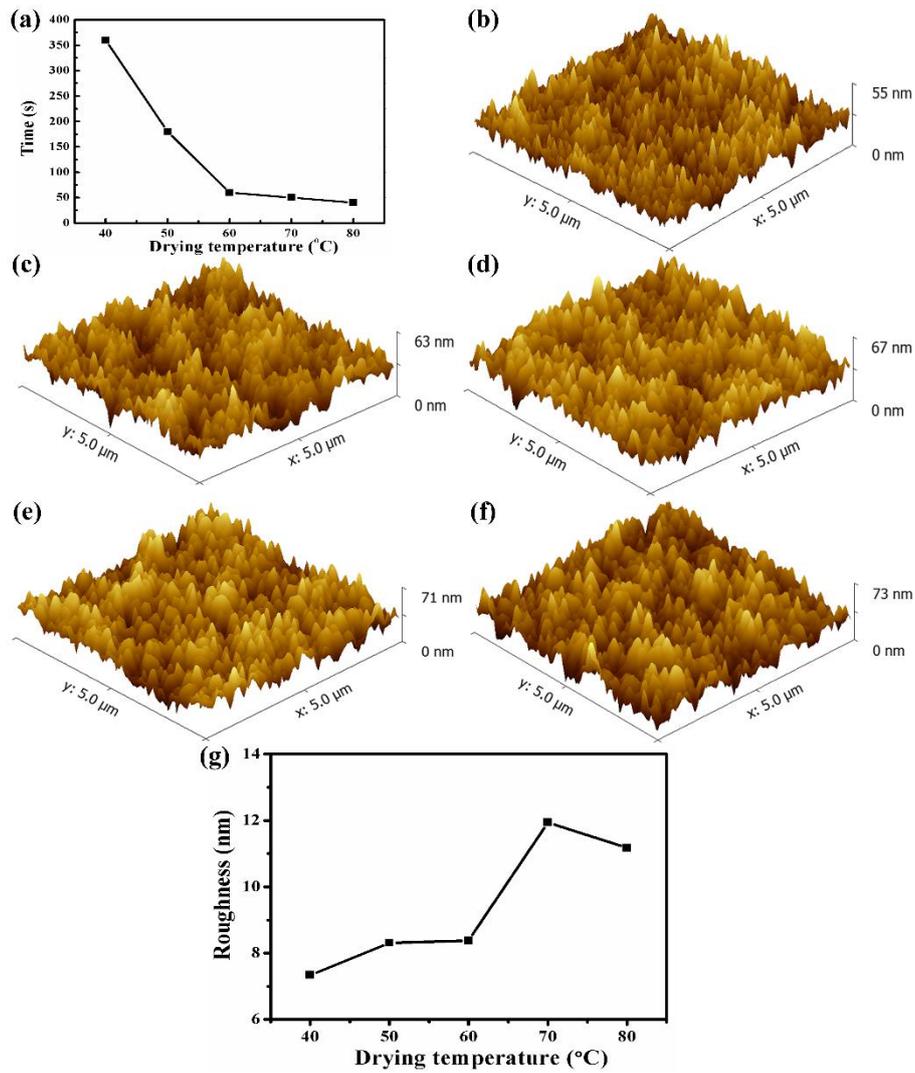

**Figure 5.** (a) The relationship between drying time and drying temperature. Three-dimensional AFM images of MAPbI$_3$ thin films at drying temperature of (b) 40 °C, (c) 50 °C, (d) 60 °C, (e) 70 °C and (f) 80 °C. (g) The relationship between surface roughness of MAPbI$_3$ thin films and drying temperature.

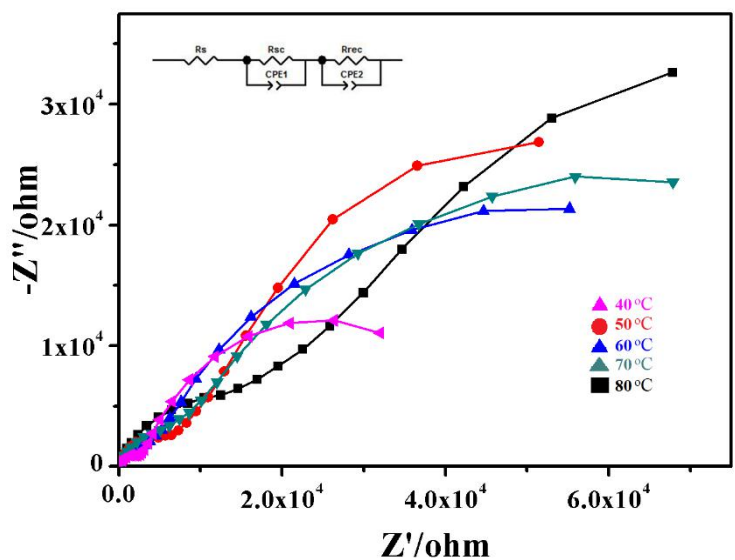

**Figure 6.** Nyquist plots of perovskite solar cells with MAPbI$_3$ films prepared at various drying temperatures in the dark with the forward bias of 600 mV.

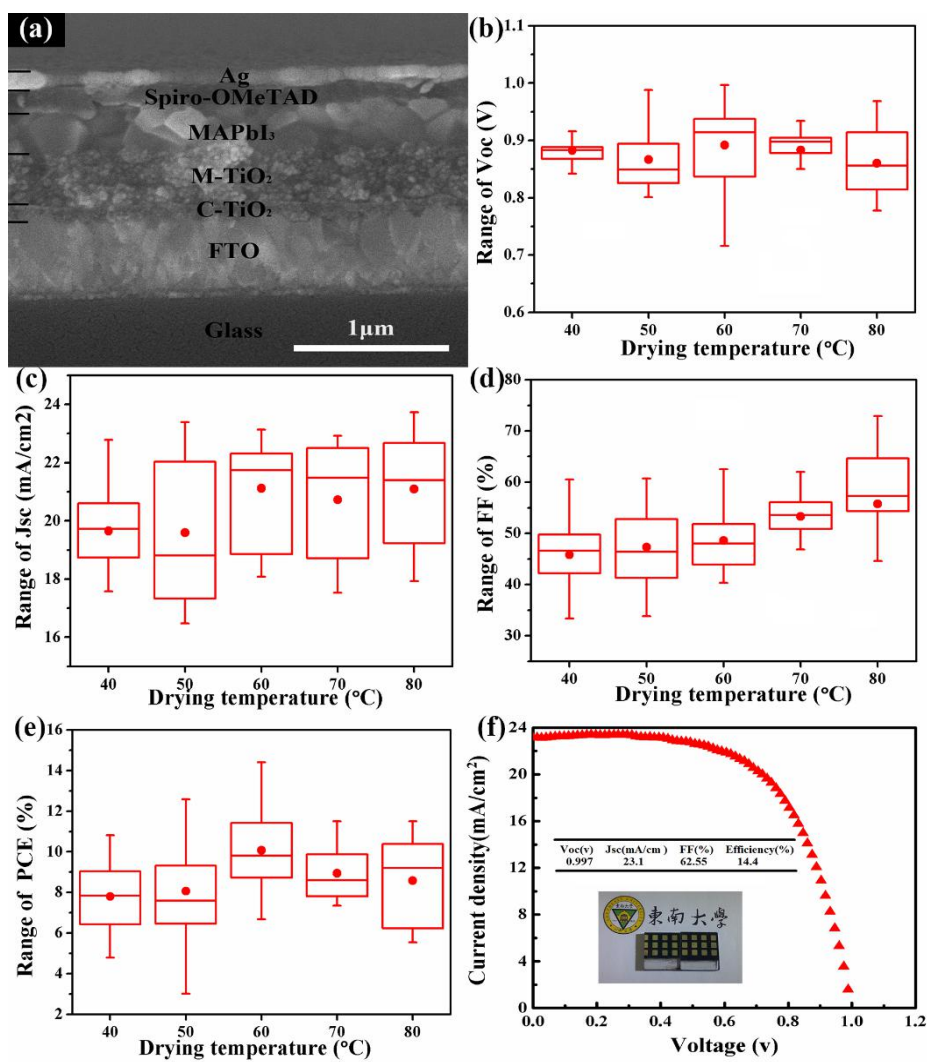

**Figure 7.** (a) Cross-sectional SEM image of perovskite solar cell with MAPbI$_3$ film prepared at drying temperature of 60 °C. Range of (b) V$_{oc}$, (c) J$_{sc}$, (d) FF and (e) PCE at the drying temperature of 40 °C, 50 °C, 60 °C, 70 °C and 80 °C. (f) J-V curves of the perovskite solar cell with MAPbI$_3$ film prepared at drying temperature of 60 °C (inset of (f) shows the photo of prepared samples).